\documentclass[preprint,showkeys, preprintnumbers,amssymb]{revtex4}

\usepackage{graphicx}
\graphicspath{{figs/}}
\begin{document}
\title{Topological effect on thermal conductivity in graphene}
\author{Jin-Wu~Jiang}
\thanks{Electronic mail: phyjj@nus.edu.sg}
	\affiliation{Department of Physics and Centre for Computational Science and Engineering,
 		     National University of Singapore, Singapore 117542, Republic of Singapore }
\author{Jian-Sheng~Wang}
	\affiliation{Department of Physics and Centre for Computational Science and Engineering,
     		     National University of Singapore, Singapore 117542, Republic of Singapore }
\author{Baowen~Li}
       \affiliation{Department of Physics and Centre for Computational Science and Engineering,
                    National University of Singapore, Singapore 117542, Republic of Singapore }
       \affiliation{NUS Graduate School for Integrative Sciences and Engineering,
                    Singapore 117456, Republic of Singapore}
\date{\today}
\begin{abstract}
The topological effect on thermal conductivity is investigated through the comparison among graphene nanoribbons, carbon nanotubes and the Mobius-like  graphene strips (MGS), by molecular dynamics simulation.  It is found that the thermal conductivity of MGS is less than one half of that of graphene nanoribbons. The underlying mechanism whereby MGS acquire such low thermal conductivity may be attributable to the enhanced phonon-phonon scattering, which is induced by the nontrivial topology of Mobius strip. Moreover by counting in the dimensions of MGS, a lower length/width ratio reduces its thermal conductivity, as the phonon-phonon scattering within might be  further elevated.
\end{abstract}


\maketitle

The phonon thermal transport in graphene has been receiving a wide range of research interest during the last decade both in theoretical development and experimental implementation. In experiment, an extremely high value of thermal conductivity of graphene (about 5000 W/mK) was measured according to the dependence of the Raman G peak frequency on the excitation laser power and the independently measured G peak temperature coefficient.\cite{Balandin} The mean free path of phonons in graphene is derived to be about 775 nm near room temperature.\cite{Ghosh} Due to the long mean free path of graphene, their physical boundary confinement  has considerable influences on thermal conductivity.\cite{Nika} Besides, the effect of Umklapp processes on the thermal conductivity of graphene was also theoretically investigated.\cite{Nika, Kong} In Ref.~\onlinecite{Kong} Kong \textit{et. al} employed first-principle approach to study the phonon thermal conductivity in both monolayer and bilayer graphenes, and found that the number of layers did not affect the in-plane thermal properties of these systems. The ballistic thermal conductance in graphene can be investigated by applying the Landauer formula.\cite{Mingo, Saito, LanJH2009, JiangJW2009} Usually, the typical behavior of the thermal conductivity in a solid is that: (1) at very low temperature where most high-frequency phonon modes are frozen, the thermal conductivity increases with temperature in a similar way as the heat capacity does; (2) The thermal conductivity reaches a maximum due to the importance of the phonon-phonon interaction when temperature is relatively high; (3) at high temperature, the thermal conductivity decreases with temperature since phonon-phonon interaction dominates in this regime. Another interesting topic is the topological electronic insulator, which origins from the intrinsic topological configuration of the system.\cite{Konig, RanY, ZhangSC, Moore} Particularly, the M\"obius-like graphene strip (MGS) has been demonstrated to be a potential one dimensional topological insulator.\cite{GuoZL} The novel electronic property of MGS with respect to its topology  gives rise to the the curiosity of the investigation of its thermal conductivity.

In this Letter, we investigate the topological effect on the thermal conductivity of graphene by molecular dynamics simulation. We comparatively study the thermal conductivity of the graphene nanoribbons, carbon nanotubes, and MGS. Our calculation shows that the topology can reduce the thermal conductivity of graphene as much as 60$\%$. Besides, we also propose that this phenomenon is induced by the stronger phonon-phonon scattering (PPS) in the MGS due to its nontrivial topology. The PPS is even stronger in MGS with a low ratio of $\gamma=length/width$, leading to a much smaller thermal conductivity of these MGS.

\begin{figure}[htpb]
  \begin{center}
    \scalebox{0.9}[0.8]{\includegraphics[width=7cm]{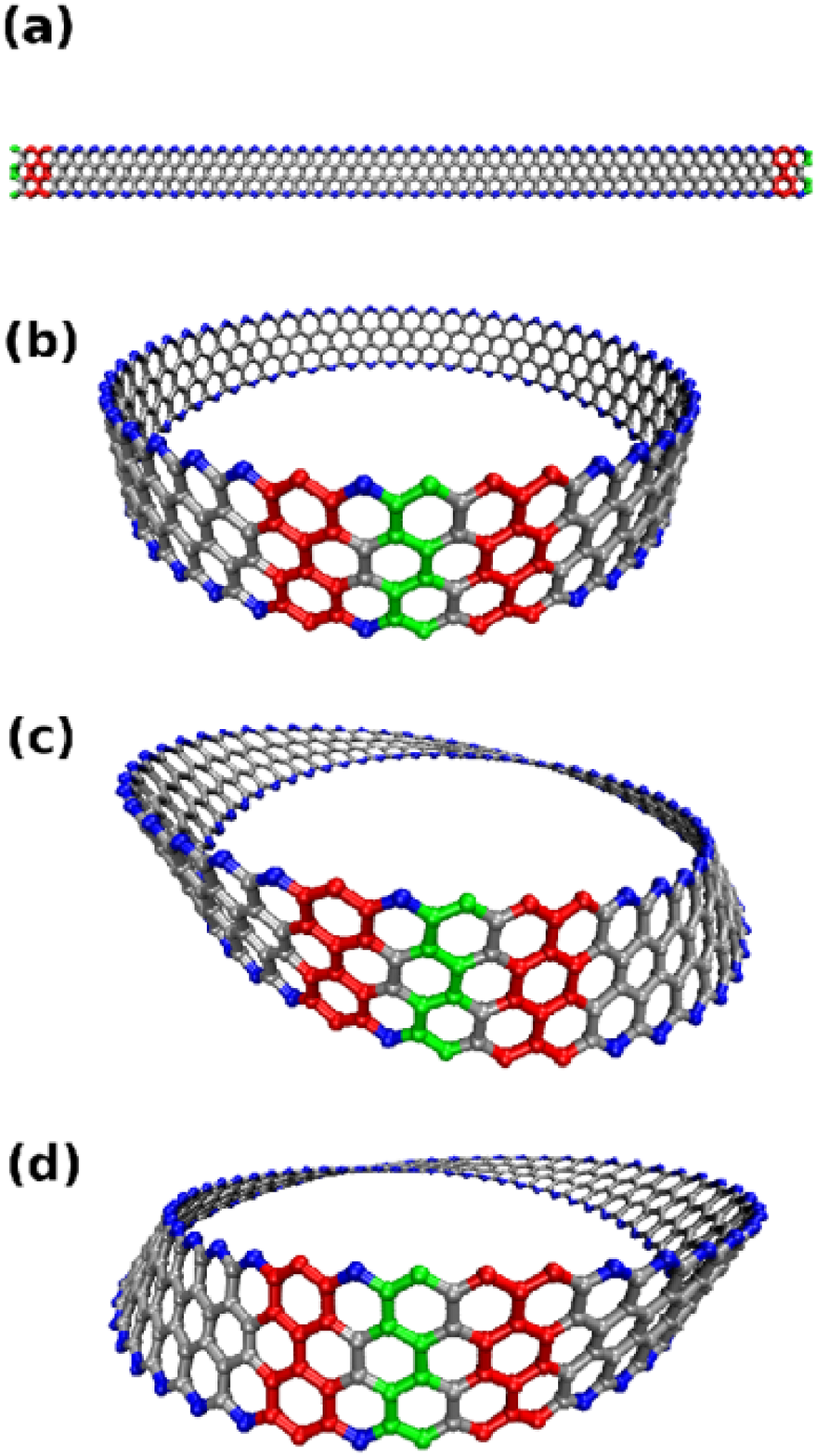}}
  \end{center}
  \caption{(Color online) Structures for different carbon-based systems: (a). graphene nanoribbon; (b). carbon nanotube; (c). MGS with chiral index $+1$; (d). MGS with chiral index $-1$.}
  \label{fig_cfg}
\end{figure}

Fig.~\ref{fig_cfg} shows the three carbon-based structures studied in this paper. All structures in the figure have zigzag boundary configuration. Fig.~\ref{fig_cfg}~(a) is an illustration of the graphene nanoribbon, where the outmost two columns (green online) on the left/right ends are fixed in the simulation. The left/right heat baths are applied to regions (red online) one column away from the left/right ends in order to avoid the temperature leaps at the boundary and to obtain a large thermal current across the system.\cite{JiangJW} For simulation convenience, the atoms on the top and bottom boundaries (blue online) are fixed. To rationalize our comparative study of the thermal conductivity of the three different structures, the same boundary conditions are applied in the carbon nanotubes and MGS. The heat baths are also applied to similar regions in all three systems. Fig.~\ref{fig_cfg}~(b) is a carbon nanotube, which is obtained by joining the two ends of the nanoribbon shown in Fig.~\ref{fig_cfg}~(a). In the nanotube, the thermal current flows in the lateral direction, not along the tube axis.  Fig.~\ref{fig_cfg}~(c) and (d) display the configuration of MGS with chiral index $\pm$. MGS can be constructed in such a way as to half-twist the nanoribbon and then join the two ends. the MGS is chiral dependent according to the direction of twisting. One of the interesting properties is that there is only one surface and one bound in MGS, on the analogy of Mobius Strip.

In our simulation, the second-generation Brenner inter-atomic potential were used\cite{Brenner}. The Newton equations of motion were integrated usin the fourth-order Runge-Kutta algorithm, where time step was chosen to be 0.5 fs. $4\times 10^{6}$ simulation steps were evolved for the system to optimize and reach thermal equilibrium state. Another $4\times 10^{6}$ simulation steps are performed to do calculation. The constant temperatures of heat baths are realized by means of N\'ose-Hoover algorithm.\cite{Nose, Hoover}

\begin{figure}[htpb]
  \begin{center}
    \scalebox{1.1}[1.1]{\includegraphics[width=8cm]{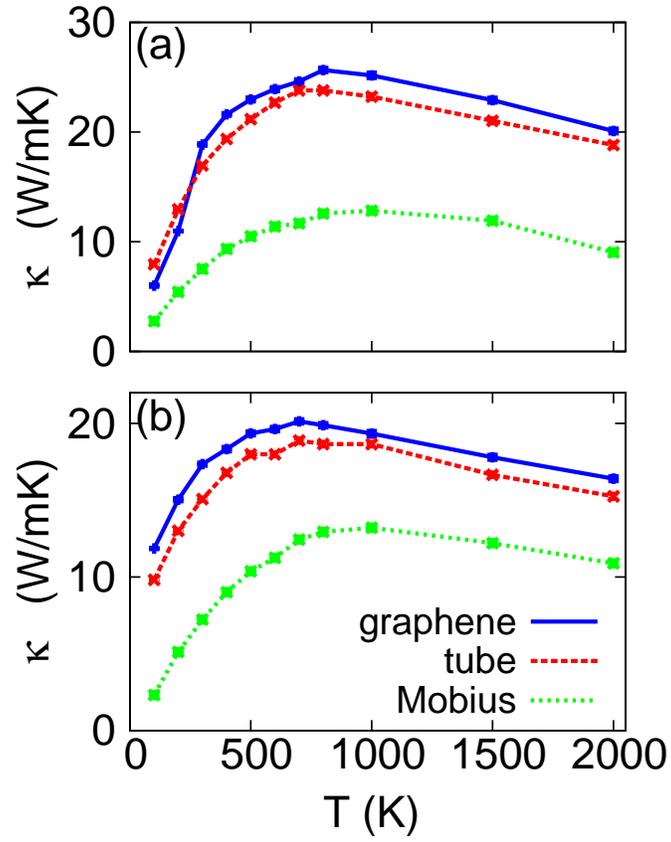}}
  \end{center}
  \caption{(Color online) Thermal conductivity at different temperatures in graphene nanoribbon, carbon nanotube and MGS: (a) with zigzag edge; (b) with armchair edge configuration.}
  \label{fig_gtm}
\end{figure}

Fig.~\ref{fig_gtm}~(a) shows the temperature dependence of thermal conductivity of the graphene nanoribbon, carbon nanotube and MGS with zigzag bondary configuration. The length of the system is 123~{\AA} and the width is 8.52~{\AA}. The quantum correction of the thermal conductivity was considered in the classical MD. It was done by introducing a quantum temperature $T_{q}$, which was calculated from equating the total ensemble energy to half of the total phonon energy. The thermal conductivity by the classical MD is mutiplied by a correction factor of $dT_{md}/dT_{q}$.\cite{Maiti} The temperatures of the three systems were restricted below $800 K$ ensuring that their thermal transports within remain in the ballistic regime on one hand, and their thermal conductivities increase with the temperature on the other. When the temperature goes beyond 800 K, the thermal transport may experience a transition from ballistic to diffusive regime. This transition temperature (800 K) is consistent with the Debye temperature in carbon covalent bonding systems.\cite{Benedict, Tohei, Falkovsky} Simulation results show that thermal thermal conductivity of the nanoribbon is highest , and that of the nanotube was slightly lower; however, the MGS showed a much lower thermal conductivity, less than half of that of the nanoribbon. Similar phenomena in the systems with armchair boundary configuration are shown in Fig.~\ref{fig_gtm}~(b) where the length and width of the system are 106.5~{\AA} and 7.38~{\AA}, respectively. To understand the mechanism of such difference, we started to study the properties of the phonon modes in these systems, which are the thermal energy carriers in their heat transport. However we found that difference between phonon modes is too small and not the dominant factor of determining the thermal conductivity of three systems (see supplement material). In turn, there should be unexposed factors responsible for the reduction of thermal conductivity in MGS.

One possible reason is the different PPS in three systems. To confirm this surmise, we analyzed the cubic nonlinear interactions:
\begin{eqnarray}
H_{n} & = & \sum_{lmn}\frac{k_{lmn}}{3}u_{l}u_{m}u_{n},
\label{eq_nonlinear}
\end{eqnarray}
where $u$ is the vibrational displacement multiplied by the square root of mass of the atom, and $k_{lmn}$ is interaction force constant matrix. This nonlinear interaction describes the three phonon scattering process, and it is obtained from the same Brenner potential by finite difference method. For convenience of comparison, we introduced a quantity, $M_{pp}$ to describe the strength of PPS,
\begin{eqnarray}
M_{pp} & = & \frac{1}{N} \sum_{lmn}|k_{lmn}|,
\label{eq_nonlinear}
\end{eqnarray}
where $N$ is the number of total atoms. Indeed, we found that the PPS in MGS is the strongest one in the three structures. The values of $M_{pp}$ in the systems discussed in Fig.~\ref{fig_gtm}~(a) are 88.6 in nanoribbon, 123.1 in nanotube, and 138.0 eV/(\AA$^{3}$u$^{1.5}$) in MGS. The largest value of $M_{pp}$ indicates the strongest PPS in MGS compared with nanoribbon and nanotube, resulting in very small thermal conductivity in MGS. These different PPS give rise to different thermal conductivities in three structures with the order as: nanoribbon $>$ nanotube $>$ MGS. This result qualitatively explained the difference of thermal conductivity. Actually, this quantity $M_{pp}$ includes the total contribution of the phonon-phonon interaction. Considering that the mean free path of the phonons is inverse proportional to the strength of phonon-phonon interaction, the quantity $M_{pp}$ should be directly related to it. The larger the $M_{pp}$, the shorter the mean free path of phonons, which lead to a lower thermal conductivity. We mentioned that although the introduced quantity $M_{pp}$ can qualitatively explain the difference of thermal conductivity in three structures, the value of $M_{pp}$ in nanotube is a little larger, which may be resulted from the roughness in the definition of $M_{pp}$. More precise method for the PPS is needed if one wants to explain quantitatively the difference of thermal conductivity in three structures.\cite{Nika} We noted that at very low temperatures where the phonon-phonon interaction was negligible, and the thermal conductivities for the three systems were almost equal to each other. This is true since the quantum correction factor $dT_{md}/dT_{q}$ is the same and almost zero in three systems. We do not consider very low temperature region, where the quantum effect dominates. This is because on one hand, it is not convincing to use the molecular dynamics to handle systems at very low temperature, although quantum correction can be done. On the other hand, the quantum corrections are the same in the studied three systems, and our paper focuses on the difference of the thermal conductivities among the three systems.

\begin{figure}[htpb]
  \begin{center}
    \scalebox{1.1}[1.1]{\includegraphics[width=8cm]{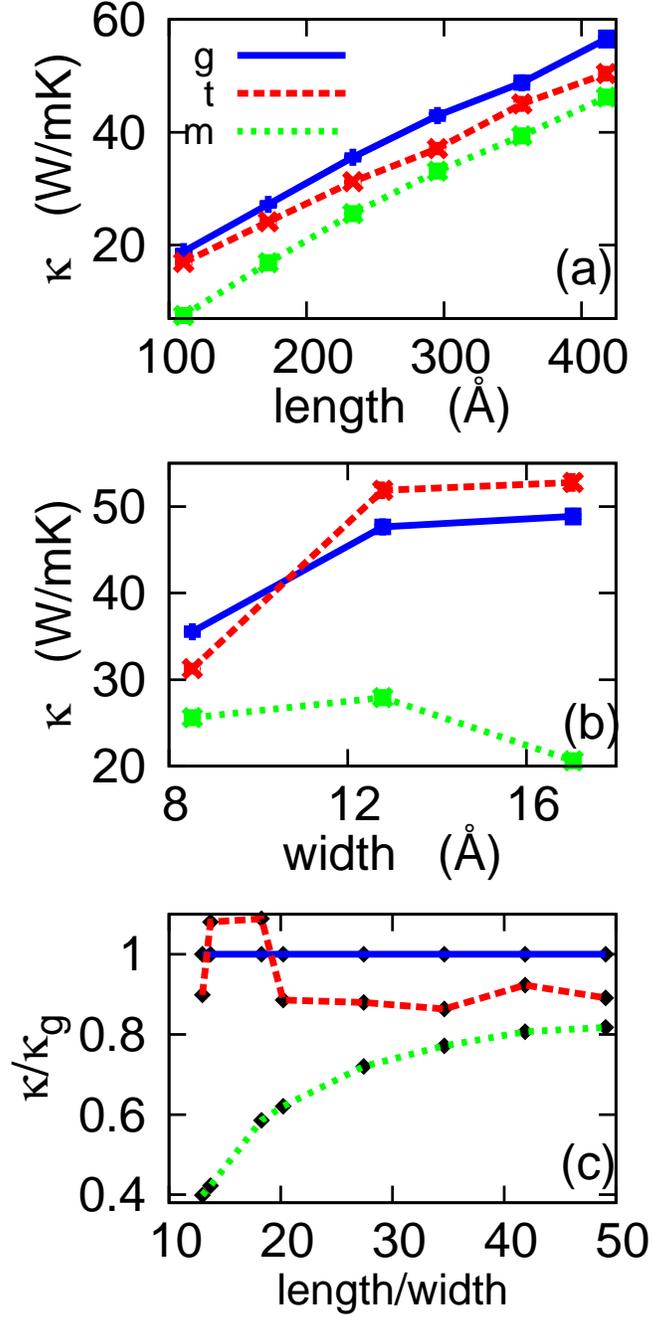}}
  \end{center}
  \caption{(Color online) Thermal conductivity at $T=300$ K in systems with zigzag edge: (a) same width (8.52~{\AA}) but different length; (b) same length (234~{\AA}) but different width. (c). Thermal conductivity (reduced by the value in graphene nanoribbon) in systems with different ratio of $length/width$. The solid line (blue online) is for graphene nanoribbon; the dashed line (red online) is for carbon nanotube; and the green dotted line is for MGS.}
  \label{fig_size}
\end{figure}

Fig.~\ref{fig_size} shows the thermal conductivity at $T=300$ K in the three systems with zigzag edge and different sizes. Fig.~\ref{fig_size}~(a) shows that the thermal conductivities of the graphene nanoribbon, the carbon nanotube and MGS increase almost linearly with their lengths, indicating the ballistic transport in these systems within these length. The difference of thermal conductivity between the MGS and the other two systems is larger in shorter systems, and this difference decreases with the system length. Fig.~\ref{fig_size}~(b) shows the thermal conductivity of the three systems with different widths. Thermal conductivity increases with the width for the graphene nanoribbon and the carbon nanotube. For the MGS with large width, its thermal conductivity shows distinct decreasing behavior. The data in (a) and (b) are shown together in (c) where the thermal conductivity in nanoribbon and nanotube are very close in whole $\gamma$ region (the deviation in small $\gamma$ region may be due to calculation error). For MGS with $\gamma=13.0$, its thermal conductivity is only about 40$\%$ of that of the nanoribbon, as a result of strong PPS with $M_{pp}=161.6$ eV/(\AA$^{3}$u$^{1.5}$) in this type of MGS.

In conclusion, by employing the molecular dynamics, we investigated the topological effect of the carbon-based systems on the thermal conductivity by the comparative study of the thermal conductivity of graphene nanoribbion, carbon nanotube and MGS. Our calculation shows that the topology of the carbon-based systems can reduce thermal conductivity up to 60$\%$. This phenomenon can be analyzed from the very different PPS in MGS compared with graphene nanoribbons and carbon nanotubes. Due to the nontrivial topological structure of MGS, the PPS within it is much stronger than that in graphene nanoribbons and carbon nanotubes, which results in much smaller thermal conductivity of MGS. The calculated $\gamma$ values of the three systems reconfirmed the aforementioned conclusion. We also found that the thermal conductivity did not depend on the chiral index of the MGS, because of the scalar property of the phonon which made it insensible to chiral index (see supplement material).

We further remark that to compare our value of thermal conductivity with the experimental results, the size effect should be considered. For example, at $T=300$ K, our calculation gave the value of the thermal conductivity $\kappa=20.0$ W/(mK) for graphene nanoribbon with the length 110~{\AA}, which is comparable with the results in Refs~\onlinecite{Lukes} with similar size. Comparing with the experimental lengths ranged in the order of $\mu$m, our result can be scale to $\kappa \times \mu$m/110~{\AA}$\approx 2000.0$ W/(mK) whic is in the same order of the thermal conductivity obtained from experimental measurements, yet still a little smaller\cite{Balandin}. One possible reason is due to the artificially fixed upper and lower boundaries, where the exited localized phonons may reduce the thermal conductivity in our systems. However, since we imposed the same boundary confinement in the three systems we investigated (nanoribbon, nanotube and MGS), the comparison of their thermal conductivities still turns out meaningful as all the calculations are based on the same benchmark.

We thank E. Cuansing and X. Wu for critical reading the manuscript. The work is supported in part by a Faculty Research Grant of R-144-000-257-112 of NUS, and Grant R-144-000-203-112 from Ministry of Education of Republic of Singapore, and Grant R-144-000-222-646 from NUS.

\end{document}